\documentstyle[referee]{mn}
\input epsf
\title{On star formation in primordial protoglobular clouds.}
\author[P. Padoan, R. Jimenez \& B. Jones]{Paolo Padoan,$^{1}$ Raul Jimenez,$^{2,3}$ 
 and  Bernard Jones$^1$\\
$^1$Theoretical Astrophysics Center, Blegdamsvej 19, DK-2100 Copenhagen, Denmark \\
$^2$NORDITA, Blegdamsvej 17, DK-2100 Copenhagen, Denmark \\
$^3$Royal Observatory Edinburgh, Blackford Hill EH9 3HJ, Edinburgh, UK}
\begin{document}
\maketitle
\begin{abstract}
Using  a new physical model for star formation (Padoan 1995) we have tested 
the possibility that globular clusters (GCs) are formed 
from primordial mass fluctuations, whose mass scale ($10^8$ - $10^9$ 
M$_{\odot}$) is selected 
out of a CDM spectrum by the mechanism of non-equilibrium formation of $H_2$. 

We show that such clouds are able to convert about 0.003 of their total mass into 
a bound system (GC) and about 0.02 into halo stars. The metal enriched gas is 
dispersed away from the GC by supernova explosions and forms the galactic disk.

These mass ratios between GCs, halo and disk 
 depend on the predicted IMF which is a consequence of the 
universal statistics of fluid turbulence. They also depend on 
the ratio of baryonic over non-baryonic mass ,$X_b$, and are 
 comparable  
with the values observed in typical spiral galaxies for $X_b \approx 0.1-0.2$. 

The computed mass and radius for a GC ( $5\times 10^5$ M$_{\odot}$ 
and 30 pc) are in good agreement with the average values in the Galaxy. 

The model predicts an exponential cut off in the 
stellar IMF below 0.1 M$_{\odot}$ in GCs and 0.6 M$_{\odot}$ in the halo. 
The quite massive star formation in primordial clouds leads to a large number 
of supernovae and to a high blue luminosity during the first two 
Gyr of the life of every galaxy.

\end{abstract}


\section{Introduction}

Globular clusters are the fossil record of galaxy formation. They are among
the oldest known objects in the Universe.
 
Since the first colour-magnitude
diagrams (CMD) for GCs were obtained by Arp, Baum and Sandage (1952) 
in the early 50's,
GCs became the natural laboratory where the theory of stellar evolution
was tested. On the other hand, the stellar evolution theory has allowed
the determination of the age of GCs (e.g., Sandage 1962, Iben \& Renzini 1984,
Vandenberg et al. 1992, Jimenez et al. 1996), that is today considered as the
best estimate for the age of the Universe.
  
Each galaxy contains hundreds of GCs, whose properties
are surprisingly similar in the Universe (Harris and Racine 1979,
Harris 1991), suggesting the
presence of a common physical mechanism in the early stages of
galaxy formation. The study of GCs
should therefore provide important clues for the development of
a theory for the origin of galaxies.
     
Sites of present day star formation do not form bound stellar systems as
massive as GCs. Therefore modelling the formation of GCs could be a
critical test for any theory of star formation.

It is known that stars are formed with an efficiency of a few percent in
giant molecular clouds of the Galactic disk 
(Duerr, Imhoff \& Lada 1982, Myers et al. 1986, Mooney and Solomon 1988), 
and bound open clusters with
an efficiency ten times smaller (Larson 1986). Therefore it is not
surprising if GCs contain only 0.1\% of the luminous mass of
the parent galaxy,
since, from the point of view of the efficiency of present day star formation,
they can only emerge from protoglobular clouds 1000 times more massive
(as in Searle 1977, Searle \& Zinn 1978, Harris \& Pudritz 1994).
	
Nevertheless models of GCs formation have identified objects of
just $10^6$ M$_{\odot}$ with protoglobular clouds. Such models
place the GC formation period either before the galaxy is formed or
during its formation. The first model for GCs as primordial
objects was proposed by Peebles \& Dicke 1968, and was later revised by
Peebles (1984) and Rosenblatt, Faber \& Blumenhal (1988). The secondary
formation scenario, in which the mass of the protoglobular clouds are 
determined by the detailed mechanism of cooling in the protogalactic cloud,  
was first proposed by Fall \& Rees (1985) and later
improved by Kang et al. (1990), Ashman (1990), Brown, Burkert \& Truran (1991), 
Brown, Burkert \& Truran (1995), Vietri \& Pesce (1995).
	 
An alternative scenario has been proposed to explain the formation of
some of the GCs in large shocks, e.g. in merging and interacting
galaxies (Ashman \& Zepf (1992), Kumai, Basu \& Fujimoto (1993)).
	
In this paper we explore the possibility that GCs are formed in large
clouds identified with primordial mass fluctuations in a cold dark matter
(CDM) spectrum. The progenitors of GCs are clouds as massive as a few 
$10^8$ M$_{\odot}$ of baryons (Searle \& Zinn 1978, Zinnecker et al. 1988  
and Larson 1990 have suggested that GCs form in the 
core of very massive clouds). Such mass-scale is selected out of the standard
CDM spectrum by the mechanism of non-equilibrium H$_{2}$ formation (see 
section 2) and, as mentioned above, it is required by the low efficiency 
of the process of star formation on large scales in order to 
form a bound stellar
system.

The protoglobular cloud is efficiently cooled to approximately 100 K 
by H$_{2}$ collisional excitation. The cooling time is short enough that  
the gas can be considered isothermal. During the dissipative collapse 
(isothermal shocks radiate away most of the kinetic energy) 
of the baryonic gas (see section 2) star formation occurs until 
supernova explosions disperse away the gas (section 4).

We suggest that this star formation process in
very massive primordial clouds can be responsible for the formation of the 
halo GCs (GCs with metallicity below 0.1 the solar value) and of a
significant fraction of the halo stars. The dispersed metal-enriched gas
is further processed in following star formation episodes.

The paper is organised in the following way. 
 In section 2 we show
how the mass-scale of protoglobular clouds is selected.
Section 3 is a brief review of the statistical model of star formation
(Padoan 1995) that is used in this work. Results are presented in sections 4
and 5. The last two 
sections of the paper contain the discussion and the conclusions.

\section{The mass of the protoglobular cloud}

If GCs have a primordial 
origin, then the mass of the typical protoglobular cloud should be 
of the order of ${M_{\rm Galaxy}}/{N_{\rm globulars}}$. On the other 
hand, it is observed that the specific frequency of GCs (Harris \& Racine 
(1979)) in different 
spiral galaxies is approximately constant, which means that there must 
 be a physical mechanism responsible for the value of the mass of the 
primordial protoglobular cloud. The value inferred from this argument is 
of a few $10^8$ M$_{\odot}$.

The rms fluctuation 
$\delta M/M$ in the mass inside a sphere of a given radius
, which is an integral over the power spectrum of density fluctuations, 
was determined by Peblees (1984) for a cold dark matter (CDM) scenario.
 
$\delta M/M$ is a shallow function of the mass scale without any particular
feature. Nevertheless, a mass scale for the first baryonic fluctuations  
that collapse can be identified. Baryonic fluctuations of mass smaller 
than this will not collapse  and fragment earlier. 

The mass-scale is selected out of the CDM spectrum as a consequence of
the mechanism of non-equilibrium formation of H$_{2}$.
In fact the collapse and fragmentation of the
gas in a gravitational potential dominated by collisionless dark matter 
particles requires energy dissipation, which becomes efficient as soon as 
a strong cooling mechanism is activated. We recognise such cooling 
mechanism in the $H_2$ collisional excitation. 

The mass-scale arises because the formation of $H_2$ in the primordial gas
(Peebles \& Dicke 1968, Hirasawa 1969, 
Hirasawa, Aizu \& Taketani 1969, Matsuda, Sato \& Takeda 1969,  
Hutchins 1976, Silk 1977, Carlberg 1981, Palla, Salpeter \& Stahler 1983, 
Lepp \& Shull 1984) occurs during the non-equilibrium recombination (cooling
time shorter than recombination time) of the H$_{2}$, that is in gas after
strong ionising shocks (Lepp \& Shull 1983, Dove \& Mandy 1986, 
Shapiro \& Kang 1987, Palla \& Zinnecker 1988). The velocities necessary
to produce such strong shocks are not present during the collapse of   
clouds of mass smaller than $10^8$ M$_{\odot}$.

Therefore efficient energy dissipation occurs only in primordial clouds
of baryonic mass larger than $10^8$ M$_{\odot}$, and these clouds
will be the first, in the CDM bottom-up hierarchy, to experience the 
fragmentation and collapse of their baryonic gas. The mass of the clouds
given by this argument is comparable to the mass of protoglobular clouds 
as inferred above from the approximately constant specific frequency
of GCs in galaxies.

We identify these primordial clouds of a few $10^8$ M$_{\odot}$ with 
protoglobular clouds.

\section{A statistical model for star formation}

A new statistical model of star formation has been recently discussed in 
a series of papers (Padoan 1995, Nordlund \& Padoan 1996, Padoan, Jones,
\& Nordlund 1996, Padoan, Nordlund, \& Jones 1996).

The model is relevant when the star forming gas is characterized by:

\begin{itemize}
\item random supersonic motions;
\item cooling time shorter than dynamical time of the random motions.
\end{itemize}

Under these conditions a complex system of interacting isothermal shocks
is present in the gas. Such flow creates a highly nonlinear density
field, with density contrasts of a few order of magnitudes. 

Since random motions are probably ubiquitous in sites of star formation,
we suggest to describe the formation of protostars as the gravitational
collapse of Jeans' masses in a density distribution shaped by random
supersonic motions. We therefore identify a protostar with one Jeans'
mass. The advantage of this description of star formation is that
a statistical description of the density field arising from random
motions in the gas can be given, and it does not depend on the detailed
phisics, but just on the rms Mach number of the flow. 

In the following subsections we briefly present the results of recent
numerical simulations of supersonic randomly forced flows, their observational
counterpart, and the derivation of the protostar mass function (MF) based
on the numerical (and observational) results.

\subsection{Numerical experiments}

Nordlund and Padoan (1996) and Padoan, Nordlund and Jones (1996)
have recently discussed the importance of supersonic flows in shaping the
density distribution in the cold interstellar medium (ISM).

They have run numerical simulations of isothermal flows randomly forced
to high Mach numbers. A description of the code and the details
of the experiments can be found in Nordlund \& Padoan (1996).

Here we just mention that the experiments consist in sloving the compressible
MHD equations in a 3-D periodic mesh. A random force is applied in 
Fourier space to small wave numbers. 

The random force produces a random flow, which is supersonic
(rms Mach number up to 10). An isothermal equation of state is used, so
that the flow soon develops a complex system of interacting shocks, with
density contrasts of a few orders of magnitudes. 

It is found that most of the mass concentrates in a small fraction of the total
volume of the simulation, with a very intermittent distribution. The probability
density function (pdf) of the density field is well approximated by a
Log-Normal distribution:

\begin{equation}
P(lnx)dlnx=\frac{1}{(2\pi\sigma^{2})^{1/2}}exp\left[-\frac{1}{2}
\left(\frac{lnx-\overline{lnx}}{\sigma}\right)^{2} \right]dlnx
\label{1}
\end{equation}
where $x$ is the relative number density:
\begin{equation}
x=n/ \overline{n}
\label{2}
\end{equation}
and the standard deviation $\sigma$ and the mean $\overline{lnx}$ are functions
of the rms Mach number of the flow, $\cal{M}$:

\begin{equation}
\overline{lnx}=-\frac{\sigma^{2}}{2}
\label{3}
\end{equation}
and
\begin{equation}
\sigma^{2}=ln(1+\frac{{\cal{M}}^{2}-1}{\beta})
\label{4}
\end{equation}
or, for the linear density:

\begin{equation}
\sigma_{linear}=\beta({\cal{M}}^{2}-1)^{0.5}
\label{5}
\end{equation}
where $\beta\approx0.5$.
Therefore the standard deviation grows linearly with the rms Mach number
of the flow.

It is also found that the power spectrum, $P(k)$, of the density distribution
is consistent with a power law:

\begin{equation}
P(k)\sim k^{-2.6}
\label{6}
\end{equation}
where $k$ is the wavenumber.

An observational counterpart of these numerical results have been recently
recognized by Padoan, Jones, \& Nordlund (1996). They show that the
infrared stellar extinction data presented by Lada et al. (1994) can
be easily interpreted if the density distribution of the absorbing
dark cloud is very intermittent. The observations are consistent with
a Log-Normal distribution. 

Padoan, Jones, \& Nordlund have shown that the standard deviation and
the power spectrum of the 3-D density distribution in the dark cloud
can be constrained by the stellar extinction measurements. They find 
values of standard deviation and spectral index that are consistent
with the numerical prediction.

In the following subsection, we show how to derive the MF of protostars,
using the random density field predicted numerically and confirmed
observationally in dark clouds.

\subsection{The derivation of the stellar IMF}

A simple way to derive the protostar MF is that of defining a protostar 
as one local Jeans' mass, so that the protostar MF is simply a Jeans' 
mass distribution. The Jeans' mass distribution is then just determined 
by the density distribution, because the gas is cooling rapidly and
therefore the temperature is uniform.

In our scenario, random supersonic motions (cascading from
larger scale) are present, and are responsible for shaping the density 
field. Strong density enhancements are due to the 
convergence of the flow, that is due to nonlinear hydrodynamical
interactions, rather than to the local gravitational potential. We suggest
therefore a description of star formation where random motions are first 
creating a complex and highly nonlinear density field (through isothermal
shocks), and gravity then
takes over, when each `local' Jeans' mass (defined with the local density)
collapses into a protostar.

Let us start from equation (\ref{1}), that is the statistic of the density,
that means the fraction of the volume occupied by any given value of the 
density. If we multiply that function times the relative density $x$, we
get the fraction of the {\it mass} occupied by any given density:

\begin{equation}
p(x)dx=xP(x)dx
\label{7}
\end{equation}

Now let us imagine to filter the density field with a filter of radius $R_{i}$,
that corrisponds to the mass scale $M_{i}$. We can sample the density 
distribution using this filter, and repeat the operation for different
filters. For any scale $M_{i}$ we get a distribution of density 
$P_{i}(M_{i},x)$.

We know, from the numerical simulations and from the observations mentioned 
above, that the Log-Normal distribution is a good model of the statistic of 
the density field. Therefore we have a Log-Normal distribution, 
$p(M_{i},x)$, for every mass scale $M_{i}$, each with its own value for 
the standard deviation, $\sigma_{i}$, and for the mean, 
$\overline {lnx_{i}}=-\sigma_{i}^2/2$.

The fraction of the total mass in collapsing structures of scale $M_{i}$ 
is the integral of the distribution $p(M_{i},x)$ along relative densities
$x > x_{J,i}$:

\begin{displaymath}
\int_{x_{J,i}}p(M_{i},x)dx
\end{displaymath}
where $x_{J,i}$ is the Jeans' density for the Mass $M_{i}$.
But of course the structure of radius $R_{i}$ can contain many local Jeans'
masses, especially if it is very dense, and therefore can fragment into
smaller objects. For this reason we must repeat the same argument for a mass
scale $M_{i+1}< M_{i}$, and subtract the collapsing mass fraction in objects
of radius $R_{i+1}$, from the previously calculated collapsing mass in
objects of radius $R_{i}$:

\begin{displaymath}
\int_{x_{J,i}}p(M_{i},x)dx-\int_{x_{J,i+1}}p(M_{i+1},x)dx
\end{displaymath}

The situation is illustrated in fig.1a, where
two filter masses, $M_{1}$ and $M_{2}$ are shown. 

Now, one should go to the
limits $M_{1}\longrightarrow M_{2}$ and $x_{J,1}\longrightarrow x_{J,2}$, 
and take the derivative 
along mass of an integral along density, 

\begin{displaymath}
\frac{\partial}{\partial M}\int_{x_{j,M}}p(M,x)dx
\end{displaymath}

where also the lower extreme of integration is subject to the derivative.

The whole procedure of extracting the protostar MF from the density
distribution can be made much easier, by assuming the distribution 
$p(M_{i},x)$ to be the same for any $M_{i}$. 
This is in fact a good approximation, because
we know from the numerical experiments and from the observations
mentioned above, that the power spectrum is $P(k)\propto k^{n}$,
where the spectral index $n=-2.6\pm0.5$. 

The standard deviation $\sigma_{i}$ is the mass variance that defines the 
power spectrum:

\begin{equation}
\sigma^2(R)=\frac{1}{2\pi^2}\int_{k}^{\infty}k^2P(k)dk
\label{8}
\end{equation} 
from which we can write:

\begin{equation}
\sigma(M)\propto M^{-\frac{n+3}{6}}
\label{9}
\end{equation}
where $n$ is the spectral index. For a spectral index $n=-2.6$ we 
obtain $\sigma(M)\propto M^{-0.07}$. Since the exponent is so small,
the distribution of density is approximately the same for any filter:

\begin{displaymath}
p(M_{i},x)dx=p(x)dx
\end{displaymath}

In other words we can say that the power spectrum is steep 
enough that the density field is almost self-similar, when looked through
different filters.

Fig.1b is the simplified version of fig.1a. Now the problem of deriving the 
mass function has become trivial, since the subtraction of the collapsing
structures of scale $M_{i+1}$, from the probability of collapse at the scale
$M_{i}$, is simply the derivative, along masses, of an integral along density,
whose integrand do not depend on mass:

\begin{displaymath}
\frac{\partial}{\partial M}\int_{x_{j,M}}p(x)dx  
\end{displaymath}

The solution is just the value of the 
integrand at $x_{J}$, times $dx_{J}/dM$. So we can see that the MF for
the protstars, $p(M)$, is just given by the transformation of the density
distribution into a Jeans' mass distribution:

\begin{equation}
p(M)dM=p(x)dx
\label{10}
\end{equation}
where $M$ is the Jeans' mass for the relative density $x$, or vice versa.

The Jeans' mass can be written as:

\begin{equation}
M=M_{J}=1M_{\odot}Bx^{-1/2}
\label{11}
\end{equation}
where:

\begin{equation}
B=4\left(\frac{T}{10 K}\right)^{3/2}\left(\frac{n}{1000 cm^{-3}}\right)^{-1/2}
\label{12}
\end{equation}
is the average Jeans' mass, that is the Jeans' mass for the average density
$x=1$.

Here we use the simplest definition of Jeans' mass: without turbulent
pressure or rotation, because the gas has just been shocked and is dissipating
its kinetic energy in a short time; without magnetic pressure, because
we will discuss the role of the magnetic field in such random flows
in subsequent papers (our numerical experiments are in fact solving the MHD
equations).

Using equations (\ref{1}), (\ref{7}), (\ref{10}), (\ref{11}), and (\ref{12}) we get the 
protostar MF:

\begin{equation}
p(M)dM=\frac{2B^2}{(2\pi\sigma^2)^{0.5}}M^{-3}exp\left[-\frac{1}{2}
\left(\frac{2lnM-A}{\sigma}\right)^2\right]dM
\label{13}
\end{equation}
where $M$ is in solar masses, and:

\begin{equation}
A=2lnB-\overline{lnx}
\label{14}
\end{equation}

One can also express the MF in average Jeans' mass, instead of
in solar masses:

\begin{displaymath}
p(\frac{M}{B})d(\frac{M}{B})=\frac{2}{(2\pi\sigma^2)^{0.5}}\left(\frac{M}{B}\right)^{-3}exp 
\left[-\frac{1}{2}\left(\frac{2ln(\frac{M}{B})-|\overline{lnx}|}{\sigma}\right)^2\right]d(\frac{M}{B})
\end{displaymath}

A log-log plot of the protostar MF is shown in fig.2. One can recognize
a long tail at large masses and an exponential cutoff at the smallest masses,
inherited from the Log-Normal distribution of density.
This shape is a good result, because most models for the origin
of the stellar IMF are not able to reproduce the cutoff at the smallest masses,
which should be present in any reasonable IMF.

The dependence of the MF on the physical parameters of the star
forming gas (average temperature, density and velocity dispersion) is
discussed in Padoan, Nordlund, \& Jones (1996), where a comparison
with the observations is also presented.

If the MF is expressed per linear mass interval (eg number of stars rather
than mass fraction), its maximum occurs at the mass:

\begin{equation}
M_{max}\approx 0.07M_{\odot}\left(\frac{n}{1000cm^{-3}}\right)^{-1/2}
\left(\frac{T}{10 K}\right)^{9/4}\left(\frac{\sigma_{v}}{2.5 km/s}\right)^{-3/2}
\label{15}
\end{equation}
(Padoan, Nordlund, Jones 1996). This is the typical stellar mass.

The result of the present derivation of the MF is that basically all the gas
turns into stars (of different masses), after the density distribution 
(\ref{1}) is established. This is a process that takes about one dynamical time
of the random motions on the large scale. Therefore the star formation 
efficiency, defined as the star formation rate per unit mass, is just the
inverse of the dynamical time of the random motions.

\section{The formation of the stellar cluster}

Since the typical mass of the protoglobular cloud is determined by the 
condition that $H_2$ can be formed during the collapse, the cooling 
time in the gas  ($10^5$  yr at a temperature of 200-300 K)  is much shorter 
than the free-fall time  (a few $10^8$ yr). Therefore the cloud is collapsing  
 at a constant temperature of about 100 K.
The collapse will be a turbulent flow with a complex system of strong 
isothermal shocks, where high density fluctuations are produced and 
protostars are born according to the model of turbulent fragmentation.
The high Reynolds number in the gas flows, the initial density inhomogeneities 
expected from the CDM scenario and the shear motions due to the tidal field 
are together responsible for the turbulence during the collapse. Moreover, 
any initial turbulence will be exponentially amplified during the collapse
 (Rotman 1991, Jacquin, Cambon \& Blin 1993).
 
In order to apply the statistical model of star formation, 
we use an average density
profile that is a power law, $\rho(r)\propto r^{-3/2}$, where $r$ is 
the distance from the center. This is the density 
profile of the inner portion of a free falling isothermal sphere (Larson 1969, 
1973, Shu 1977).

Using this density profile, the temperature of 100 K and the 
turbulent velocity dispersion equal to half of the free fall velocity, 
we can calculate the IMF and the 
star formation rate 
 as a function of the distance from the center of the cloud during the  
 collapse of this. 
 
Since the density is higher in the center than in the outer parts, the star 
formation rate is increasing towards the center. This means that at 
some point during the collapse a bound stellar system 
could be formed around the center, while the 
remaining gas would be dispersed by supernova explosions. The main 
aim of this work is to test if that bound system looks like a GC and 
if the scenario is consistent with observational constraints related to the 
formation of the Galaxy.

\subsection{Formation of a bound system}

The star formation process continues as long as the collapse is taking place,
 that is until the energy injected in the system by 
 supernova explosions becomes larger than the gravitational energy of the 
  cloud. When this happens, if at all, a bound stellar system will 
 be left only if the total mass of the stars is comparable or larger than 
 the mass of the dispersed gas (Lada, Margulis \& Dearborn 1984).

 \begin{figure}
\centering
\leavevmode
\epsfxsize=1.0
\columnwidth
\epsfbox{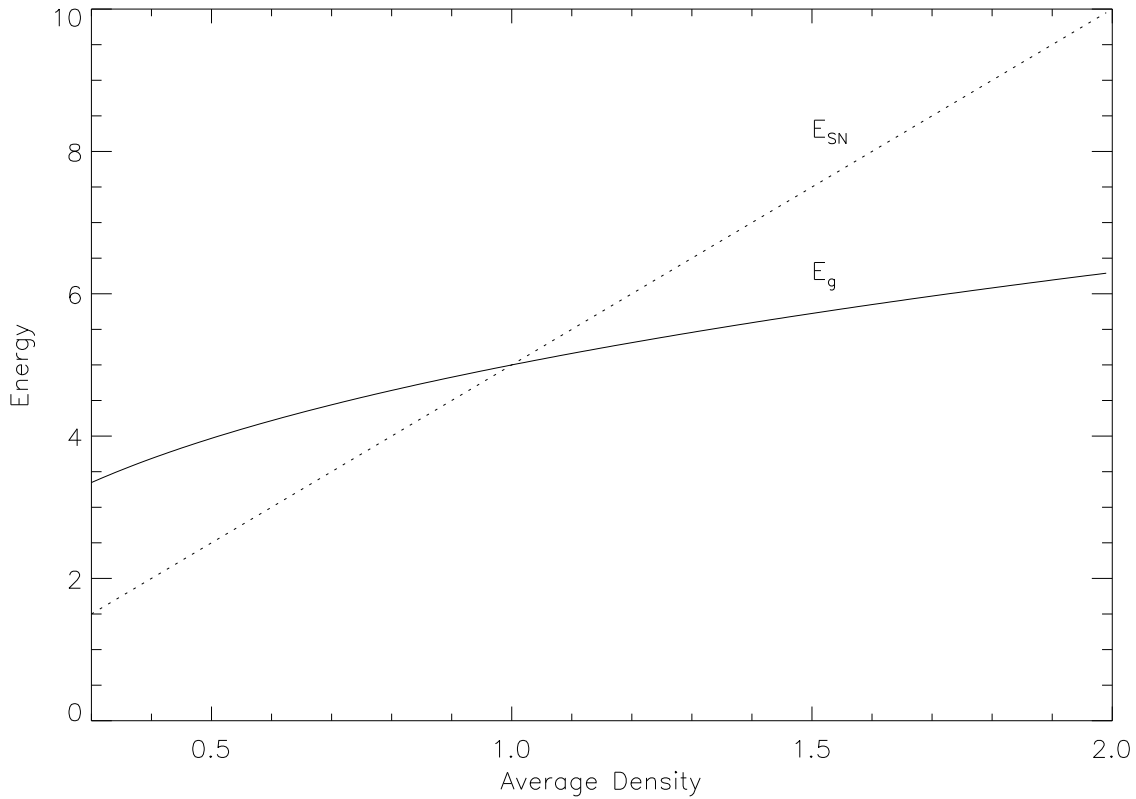}
 \caption{Total gravitational energy of the protoglobular cloud (continuous line) and kinetic energy input from supernovae (dashed line). The energy is 
 in arbitrary units and the density in $cm^{-3}$. The cloud is
 destroyed by the supernovae when it has reached an average density of about 
 $1 cm^{-3}$.}
 \end{figure}

We compute the star formation rate and the IMF during the collapse - 
i.e, at different densities. At any given density, we compare the 
total gravitational energy of the cloud with the kinetic energy 
injected into the gas motions by explosions of stars more massive than 
8 M$_{\odot}$. Every supernova is supposed to liberate an energy of 
$10^{51}$ ergs that is turned into kinetic energy of the surrounding gas with  
an efficiency of 1\% (examples of efficiency estimates are given in 
Spitzer 1978, Brown, Burkert \& Truran 1995).

In Fig. 1 we show that the total energy of the gaseous cloud becomes 
positive when the average cloud density is equal to 1.0 cm$^{-3}$. At this 
density the collapse is stopped and the gas is dispersed away from the 
stars, marking the end of the star formation process.

In order to check if a bound stellar system is formed at all, we compute 
the star formation efficiency (mass star/total mass) at different distances
from the center of the cloud; the bound stellar system is formed wherever 
the efficiency is close to 100\%, in Fig. 2 the star formation efficiency 
is plotted versus a radial coordinate (mass fraction). The efficiency falls 
below 50\% at a mass fraction of about 0.004, that corresponds to a radius 
of 50 pc. 
  
Therefore {\it our star formation model applied to a primordial cloud 
of $2.5\times10^8$ M$_{\odot}$ predicts the formation of a bound stellar system
with a mass of  $5\times10^5$ M$_{\odot}$ and a radius of 30 pc  
}.

The integration of the function plotted in  Fig. 2 shows that approximately 
0.002 of the total mass of 
the cloud has turned into a bound stellar system and 0.02 into 
stars located outside 
the bound system, that  
will evaporate and eventually form the halo of the Galaxy (without  
including the bulge component).

\subsection{Initial mass function}

The IMF and its typical mass are given by the formula 1 and 6 (section 3). 
The IMF for GCs is plotted in Fig 3 (dashed line). The IMF has a single maximum, an 
exponential cut off for smaller masses and a long intermittent tail for 
larger masses. In the figure we plot with the continuous line 
the exponent ($x(m)$) of the power law 
IMF that approximates the predicted IMF. The slope for masses below  
 1 M$_{\odot}$ is always smaller than the Salpeter, $x=1.35$,
and  the maximum is located at 0.1 M$_{\odot}$. The slope becomes equal 
to the Salpeter one around 2 M$_{\odot}$ (not shown in Fig 3).

Therefore our model predicts that {\it deep luminosity functions observed 
in GCs should be characterised by a cut off at luminosities corresponding
to masses of about 0.1 M$_{\odot}$ preceded by a progressive flattening of 
the distribution}.

\section{Globular clusters and the Galaxy}

The picture of GC formation proposed in the present work implies the 
following scenario for galaxy formation:

\begin{itemize}

\item The protogalaxy is an ensemble of a  
few hundred primordial clouds of about $10^8$ M$_{\odot}$ 
that are in self collapse.

\item One GC is formed inside the nucleus of each cloud.

\item The remaining part of the cloud forms the halo stars that evaporate 
from the cloud.

\item The cloud gas that is expelled out by the supernova explosions falls 
dissipatively into a disk system.

\end{itemize}

As we mentioned above, most of the stars formed outside the bound system 
are evaporated from the cloud. The total mass in this stellar component 
is about 10 times larger than the mass of the GC. Moreover, the 
total mass of the dispersed gas is around 500 times the mass of the GC.

Therefore our model predicts reasonable mass ratios 
between GCs and galaxy masses (Harris \& Racine 1979) and between GCs and 
the stars in the external halo (Woltjer 1975, Bahcall \& Soneira 1982, 
Harris \& Racine 1979). 
Moreover,   
it also predicts the IMF for halo stars and the initial metallicity of 
the galactic disk. The halo IMF is shown in Fig 4.

\begin{figure}
\centering
\leavevmode
\epsfxsize=1.0
\columnwidth
\epsfbox{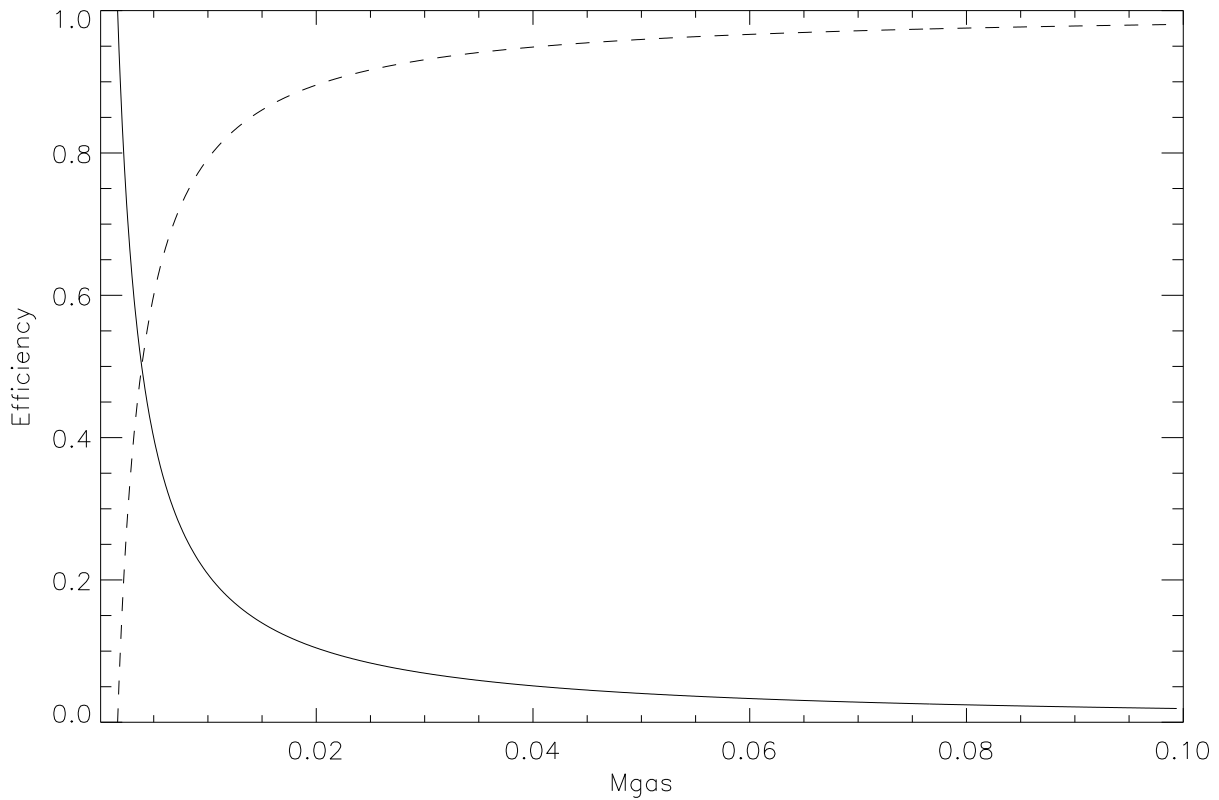}
\caption{Star formation efficiency (mass in stars over initial gaseous mass)
versus mass fraction (0.0 in the center, 1.0 at the largest radius) at the
end of the star formation process. The dashed line shows the fraction of left
over gas at any position from the center.}
\end{figure}

The most probable mass of halo stars is 6 times more massive than the 
one in GCs. This is due to the fact that halo stars are formed at a lower 
density in more external regions of the primordial cloud compared to
GC stars. So we predict that {\it deep stellar counts of halo stars 
will show a cut off in the IMF below 0.6 M$_{\odot}$}.

The gas that falls into the disk system is previously 
metal-enriched by the 
star formation in the protoglobular clouds. Therefore 
 we expect to find virtually {\it no disk 
stars with metallicity lower than [O/H]$=-1$}.

\section{Discussion}

In this paper we have presented a model for the formation of the halo
population of stars (bulge excluded) and GCs (metallicity below 0.1 the solar value).
The visible mass of such population in the Galaxy is a few percent of the
total Galactic mass for the halo stars and a few per thousands for the GCs. 
These mass ratios are reproduced by the present model for reasonable 
values of the non-baryonic mass of the Universe.

Searle \& Zinn (1978) and Zinn (1985) have proved the existence of two distinct 
populations in the system of Galactic GCs. The two populations differ in
metallicity and kinematic properties. The halo GCs show no metallicity
gradient outside 8 Kpc from the Galactic center while the disk GCs
show a metallicity gradient and a minimum metallicity [Fe/H]=-1.

Several authors have proposed distinct mechanisms for the origin of the
two GCs populations (Searle \& Zinn 1978, Zinn 1985, Rosenblatt, Faber \& 
Blumenthal 1988, van den Bergh 1993), while no attempt has been made here 
to model the formation of the disk GCs. 

The halo GCs have been explained as the result of the star formation process
in primordial mass fluctuations in a CDM spectrum; halo stars are also
formed in this way. The typical mass of primordial protoglobular clouds is
explained by assuming that the onset of H$_{2}$ cooling is essential 
for the collapse and fragmentation of the baryonic gas in gravitational 
potentials 
dominated by collisionless dark matter. In fact, it is the short cooling time
in the gas that makes the shocks isothermal, so that 
purely baryonic strong density enhancement can be achieved, and the collapse
and fragmentation of the baryonic gas can proceed.

\begin{figure}
\centering
\leavevmode
\epsfxsize=1.0
\columnwidth
\epsfbox{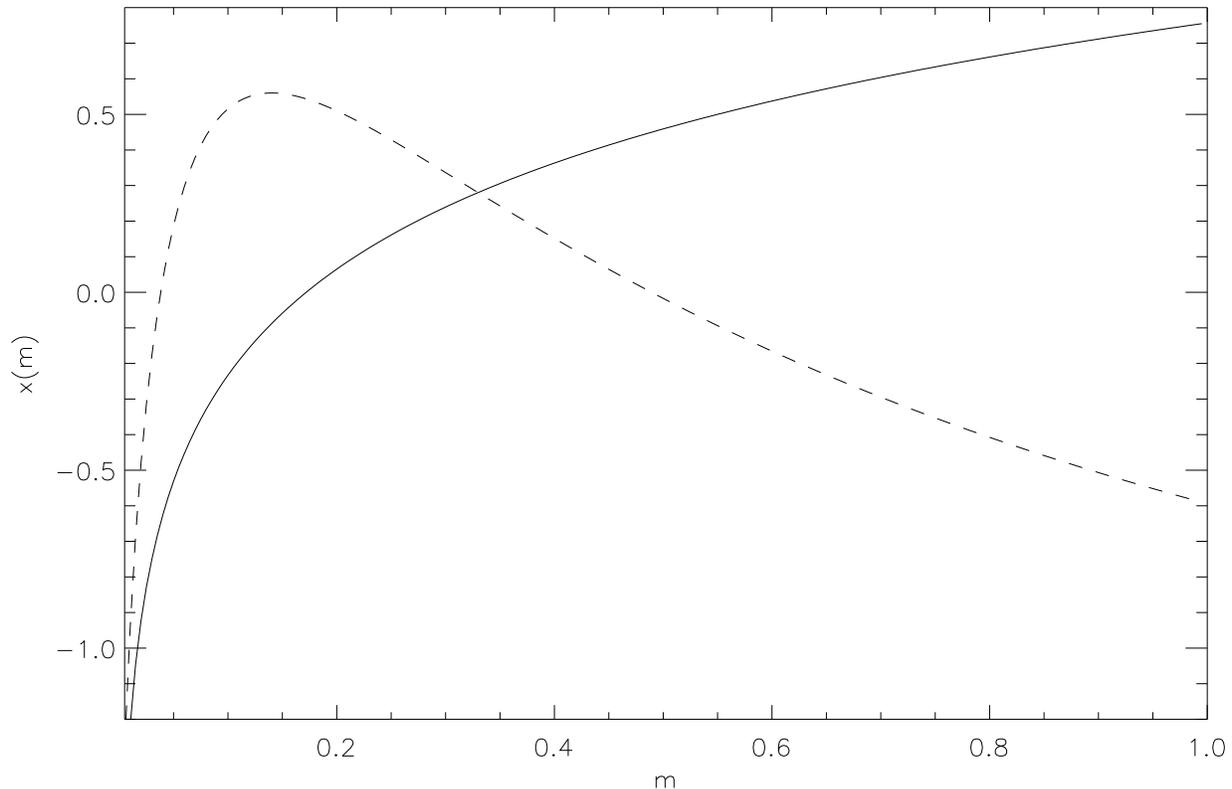}
\caption{IMF in GCs. The dashed line shows the IMF in arbitrary units: the
maximum is at 0.1 m$_{\odot}$. The continuous line gives the exponent ($x(m)$)
of the
power law IMF that approximates that segment of the predicted IMF. The
slope for masses below 1 M$_{\odot}$ is always smaller than the Salpeter
x=1.35.}
\end{figure}

We have found that the ratio of baryonic to non-baryonic mass, $X_{b}$  
plays an important role in the formation of the combined system of halo 
GCs and halo stars. If $X_{b}\stackrel{<}{\sim}0.1$ 
too massive GCs are formed, in   
contradiction with the observations. On the other hand the formation of 
the estimated value of the mass of 
halo stars requires $X_{b}\stackrel{<}{\sim}0.2$. 

We can conclude that the primordial origin of halo GCs and halo stars in 
the CDM scenario is feasible only if $X_{b}=0.1-0.2$, that is the density 
parameter $\Omega=0.2-0.5$, if the baryonic density parameter 
$\Omega_{b}=0.05$ (Kolb \& Turner 1990).

\subsection{The initial mass function}

Our statistical model of star formation is based on a fragmentation
mechanism controlled by the statistics of fluid turbulence. The universal 
character of the statistical properties of turbulence generate a 
functional form of the stellar IMF which does not depend on environmental
physical parameters, even though the typical stellar mass does. 

The GCs IMF has its low mass cutoff at 0.1 M$_{\odot}$. The slope of the 
IMF (x=1 at 0.8 M$_{\odot}$) is smaller than the Salpeter one (x=1.35), and
becomes comparable with it at about 2 M$_{\odot}$.

There have been attempts to determine the IMF in GCs. Fahlman et al. (1989), 
Richer et al. (1991) and Piotto \& Ortolani (1991) find very
steep IMFs. Nevertheless these
determinations are affected by several uncertainties, the most important 
being the unknown mass-luminosity relation. 
Moreover the measured IMFs
span only a very limited range in masses (0.2-0.7 M$_{\odot}$) in the best 
case, making
any extrapolation for other masses very uncertain.

Recent HST observations find 
a cutoff in the 
IMF at about 0.1 M$_{\odot}$, in good agreement with our 
prediction. 

Models of GCs evaporation and core collapse (Hut \& Djorgovsky (1992)) are
consistent with observations if the average stellar mass is assumed 
to be 0.3 M$_{\odot}$, consistent with our model.

The expected non-baryonic dark matter inside the GC is only
a small fraction of the GC mass because the high velocity (about 100 km/s) 
of 
the dark matter particles prevents them from falling into the baryonic 
central density condensation. As for the baryonic dark matter, our
IMF contains a negligible number of brown dwarfs and jupiterian objects. 
Therefore the total dark matter content of GCs is expected to be small, 
in agreement with 
the observations.

The predicted stellar masses in the halo are 6 times larger than in the
GCs. This is due to the smaller average density of the gas where halo
stars are formed (the maximum in the halo IMF corresponds to a 
density of 250 $cm^{-3}$ and in the GC of 10000 $cm^{-3}$). 
The halo stellar population is also more massive than
the solar neighbourhood one, suggesting that the star formation process
in the early protogalaxy is characterised by the presence of a large number
of supernovae, compared with the subsequent star formation in the disk. 
Therefore the contribution to brown dwarfs and jupiters from the halo 
stars is also negligible.

The halo IMF implies that in the evolution of every galaxy there
will be a high blue luminosity during the first 2 Gyr. This event should 
be included in models of galactic luminosity evolution to be used for the 
interpretation of deep galaxy counts.

We predict a halo IMF that is less steep than the  
Salpeter's one below 2 M$_{\odot}$.
Richer \& Fahlman (1992) observed low mass stars down to 0.14 M$_{\odot}$ in 
the halo and suggested that the slope of the IMF is instead steeper than the 
Salpeter's one. On the other hand, more recent HST observations by Bahcall et 
al. 1994 of high latitude fields found fewer faint red stars. 

\subsection{Metallicity}

Halo GCs in large galaxies show a dispersion in metallicity that is much 
larger than the internal one (Harris and Racine (1979)); this fact is 
sometimes used as an argument against the primordial origin of GCs, 
because it is believed that self-enrichment in the protoglobular cloud 
should lead to a heterogeneous metallicity distribution inside the GC. 

Nevertheless in our picture the stars responsible for the enrichment 
are formed during the dissipative collapse of the baryonic cloud, which 
takes a few $10^8$ years. This time scale is long enough to allow 
for efficient mixing if only the central region of the cloud, 
where the GC will be formed, is considered. 
The bulk of GC stars are formed in a period of only 
$10^7$ years (because of the stellar feedback that destroys the cloud),  
when the central average density has become large enough, and 
therefore such stars cannot be responsible for any further self-enrichment. 
On the other hand the efficient mixing for the whole cloud requires a 
longer time than for the nucleus, so that some stars will be formed  
outside the GC with very low metallicity. In fact the metallicity
distribution of halo stars is known to extend to practically zero
metallicity (Beers \& Sommer-Larsen (1995)).

\begin{figure}
\centering
\leavevmode
\epsfxsize=1.0
\columnwidth
\epsfbox{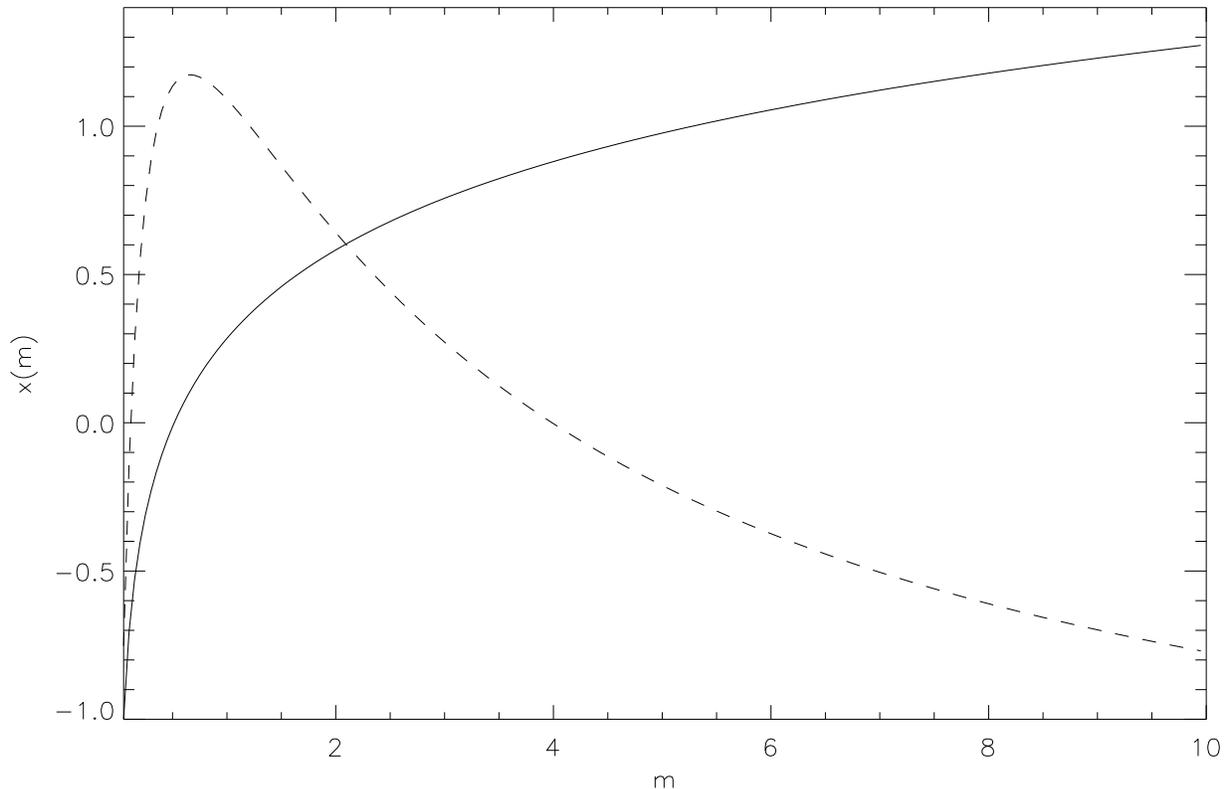}
\caption{Halo IMF. As in fig. 3. The maximum is now at 0.6 M$_{\odot}$ and
the slope equals the Salpeter x=1.35 only for masses larger than
10 M$_{\odot}$.}
\end{figure}

Only a few supernovae are required to enrich the primordial cloud and 
achieve the value for the metallicity observed in halo GCs.  
The small number of supernovae easily explains the range in metallicity  
observed in halo GCs, as due to statistical fluctuations in the number of 
supernovae events, before the final star formation burst.

Our scenario for GC formation is not only consistent with metallicity 
determinations in GCs but also with the observed lower limit for 
the disk metallicity: 
there are virtually no disk stars found with metallicity lower than 
[O/H]$=-1.0$ (van den Bergh 1962, Schmidt 1963, Beers \& Sommer-Larsen 1995).   
We calculate the oxygen yield from the star formation in the protoglobular 
cloud, and find that the metallicity of the gas, that is dispersed from the 
cloud and falls into the disk, is about [O/H]$=-1.0$. 
So we do not expect any disk 
star to be formed with  metallicity lower than this value, 
in agreement with the observations.

The observed abundances of oxygen over iron in halo stars 
(Matteucci \& Greggio 1986, Wheeler, Sneden \& Truran 1989, McWilliam 
\& Rich 1994), and the fact that the median halo 
stars metallicity (Norris 1986) is 
about the same as the median metallicity in GCs, 
suggest that halo 
stars and GCs belong to the same star formation process as stated in our 
model.

\section{Conclusions}

The main conclusions of this paper are the following:

\begin{itemize}

\item GCs can be formed out of the nucleus of a  primordial cloud of 
 a few $10^8$ M$_{\odot}$ of baryons. 

\item The mechanism of $H_2$ formation selects this typical mass scale out
of a CDM spectrum.

\item Most halo stars in the galaxy (bulge excluded) can be formed together 
with the (halo) GCs in the same clouds; the halo and the GCs masses are 
sensitive to the ratio of baryonic over non baryonic matter, X$_{b}$, in 
the cloud. 

\item A ratio $0.1<X_{b}<0.5$ produces realistic values for the mass of GCs;
if $X_{b}<0.1$ too large GCs are formed.

\item A ratio $X_{b}<0.2$ is needed in order to originate most of the 
halo stars in the protoglobular clouds during the formation of GCs.

\item The gas that is blown away from the disrupted 
protoglobular cloud by supernova explosions is used in the following
star formation episodes; therefore the initial 
disk metallicity (minimum stellar metallicity in the disk) is [O/Fe]$=-1$.

\item The IMF in GCs has a cut off below 0.1 M$_{\odot}$ while the 
halo IMF has the cut off below 0.6 M$_{\odot}$.

\item During the first 2 Gyr of its life each galaxy has higher blue   
luminosity than normal disks.
This is due to the fact that star formation  (halo stars) 
in the 
protoglobular cloud produces a significant number of massive stars, 
which also means an epoch of large number of supernova explosions.

\end{itemize}

We have shown that this scenario is consistent with several observations.

The star formation rate and the stellar IMF are essential ingredients
of the model. In order to predict them we have used a new physical
model of star formation. In this model the functional form of the IMF 
is shown to be universal because it is based on the statistics of 
turbulence which are universal properties of flows.

Future work should concentrate in the attempt of simulating the collapse 
of CDM primordial protoglobular clouds. Cosmological simulations on 
such small scales are at the moment beyond the capabilities of available 
computer resources, since they require a very large dynamical range.

\section*{Acknowledgements}

We thank S.M. Fall and A. Kashlinsky for fruitful comments 
and discussions during the 
elaboration of this manuscript. We also enjoyed discussions with 
C. Flynn, C. Lacey \& B. Pagel.
We are thankful to the referee, Dr. Joe Silk, for helpful comments on 
the paper.
 
 We are grateful 
to Michele Cavigioli for providing special computer facilities in Copenhagen.

The work has been partly supported by the Danish National Research Foundation 
through its support for the establishment of the Theoretical Astrophysics Center.

\end{document}